\documentclass[5p,twocolumn,times,number]{elsarticle}

\usepackage{graphicx}
\usepackage{amsmath}  

\usepackage{lineno}

\begin{document}

\begin{frontmatter}

\title{TPC-like readout for thermal neutron detection using a GEM-detector }

\author[add1]{B.~Flierl\corref{cor}}
\ead{bernhard.flierl@physik.uni-muenchen.de}
\author[add1]{R.~Hertenberger}
\author[add1]{O.~Biebel}
\author[add2]{K.~Zeitelhack}

\cortext[cor]{Corresponding author}

\address[add1]{Ludwig-Maximilians Universit\"{a}t, M\"{u}nchen, Germany}
\address[add2]{Forschungs-Neutronenquelle Heinz Maier-Leibnitz, Garching, Germany}

\begin{abstract}
Spatial resolution of less than 200 $\mu$m is challenging for thermal neutron detection. A novel readout scheme based on the time-projection-chamber (TPC) concept is used in a gaseous electron multiplier (GEM) detector \cite{Sauli:1997qp}. Thermal neutrons are captured in a single 2 $\mu$m thick Boron-10 converter cathode and secondary Helium and Lithium ions are produced with a combined energy of 2.8\,MeV. These ions have sufficient energy to form straight tracks of several mm length.
With a time resolving 2-dimensional readout of 400 $\mu$m pitch in both directions, based on APV25 chips, the ions are tracked and their respective origin in the cathode converter foil is reconstructed.
Using an Ar-CO2 93:7\% gas mixture, a resolution of {100 $\mu$m} (FWHM {235 $\mu$m}) has been observed with a triple GEM-detector setup at the Garching neutron source (FRMII) for neutrons of 4.7 \AA.

\end{abstract}


\end{frontmatter}

\section{Introduction}
Thermal neutron detection with spatial resolution 
around or below 0.2\,mm will become mandatory for 
modern thermal neutron crystallography or fast imaging detectors \cite{kirstein2014neutron}.

Gaseous micro-pattern detectors like
GEM (gaseous electron multiplier) or Micromegas 
have proven to reach spatial resolutions well 
below 100 $\mu$m \cite{MM_res}, 
depending on the fineness of their segmentation. Here a self calibrating TPC-like analysis for this kind of detector is presented with a GEM-detector.

\section{TPC-like Analysis of Ion Tracks}
A triple GEM-detector with segmented 2D-readout of 0.4 mm pitch was used in combination with a $^{10}$B converter cathode to detect thermal neutrons. Thermal neutron capture in $^{10}$B has a high cross section and produces a $^7$Li and a $^4$He ion back-to-back with a combined energy of 2.8 MeV.
\begin{align*}
^{10}B + n &\rightarrow ^7 Li + ^4 He + \gamma + Q(2.79 MeV)  (94\%) \\
^{10}B + n &\rightarrow ^7 Li + ^4 He  + Q(2.79 MeV) (6 \%)
\end{align*}  

%and thus the 
%interaction point of the thermal neutron in the 
%converter foil 
The heavy He or Li ion 
create a straight line of ionized electrons
in the drift-volume of a micro-pattern detector. The full track information  
can be reconstructed from the electron drift time 
measured as a function of 
the responding readout segments. 
This allows for reconstruction of the 
neutron interaction point in the converter cathode.

The APV front-end-electronics samples the signal 
on every strip in steps of 25 ns. With typical rise times of 150 ns  
the beginning of a signal, and thus the drift time of the respective electrons, 
can be determined with a fit to the rise of the charge distributions.

  \begin{figure}[h]           
 \centering
                \includegraphics[width=0.35\textwidth]{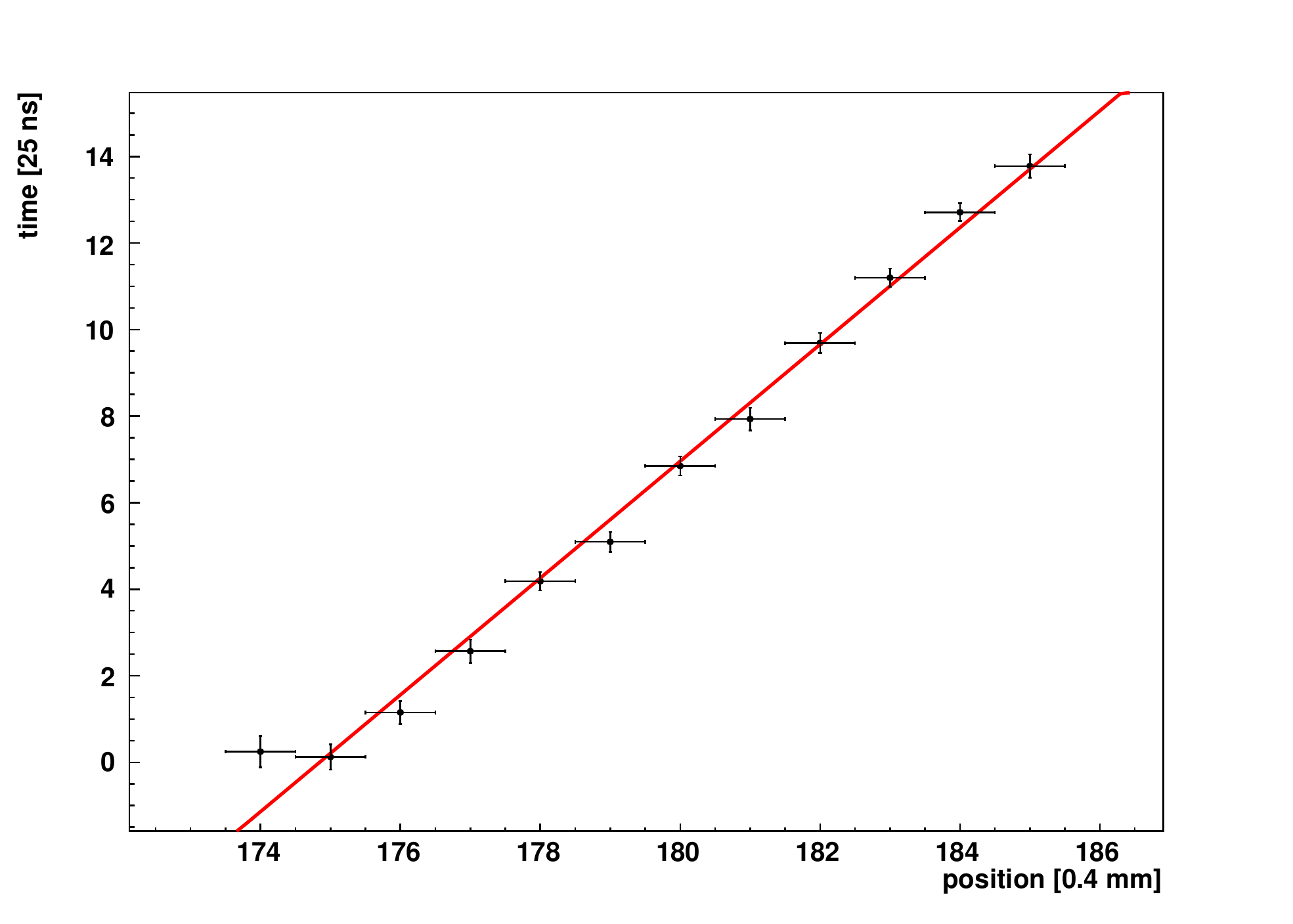}
                \caption{The starting time of the signal is determined for every strip by fit 
		to the rise time of the respective charge signal. 
		Direction and orientation of the track are determined by a linear fit to the start points}
                \label{fig:raw1}

\end{figure}
From a fit to the starting times as a function of hit strips the inclination angle of the track can be reconstructed.

\section{Determination of spatial resolution}

As conversion cathode a 2 $\mu$m layer of $^{10}$B evaporated on an aluminum supporting sheet has been used. 
It was mounted to a triple GEM detector with two-dimensional strip readout, based on crossed strips with a pitch of 400 $\mu$m in both directions. The readout anode was delivered by CERN. X and Y directions are covered by 256 strips each with a copper strip width of 80  $\mu$m for the upper layer and 320  $\mu$m for the lower layer. All the insulator material but the direct insulation between X and Y copper strips have been etched off, such that the different copper width guarantees similar pulse-height in both planes.
The readout is based on APV-25 front-end boards \cite{APV} and the SRS data acquisition system  \cite{SRS} and was triggered by the signal on the lower side of the last GEM-foil, which is in principle the inverted signal of the readout plane.
The drift gas, Ar-CO$_2$ of 93:7 Vol.\% at atmospheric pressure, defines in combination with the thickness of the absorber the maximum range of the $^4$He ions to 12 mm. The width of the drift gap of 6\,mm was chosen to decrease the impact of the Bragg-peak in the track reconstruction and to assure that only tracks with large inclination angle stop in the active volume. The distances between GEM-foils and the distance between last GEM-foil and the readout anode was 2 mm each.
For optimized spatial resolution the drift velocity of the electrons was chosen quite low to 11 $\mu$m/ns, by adapting the electric field in the drift region to the time resolution of the APV25 chips.

\begin{figure}
\centering

\includegraphics[width=\linewidth]{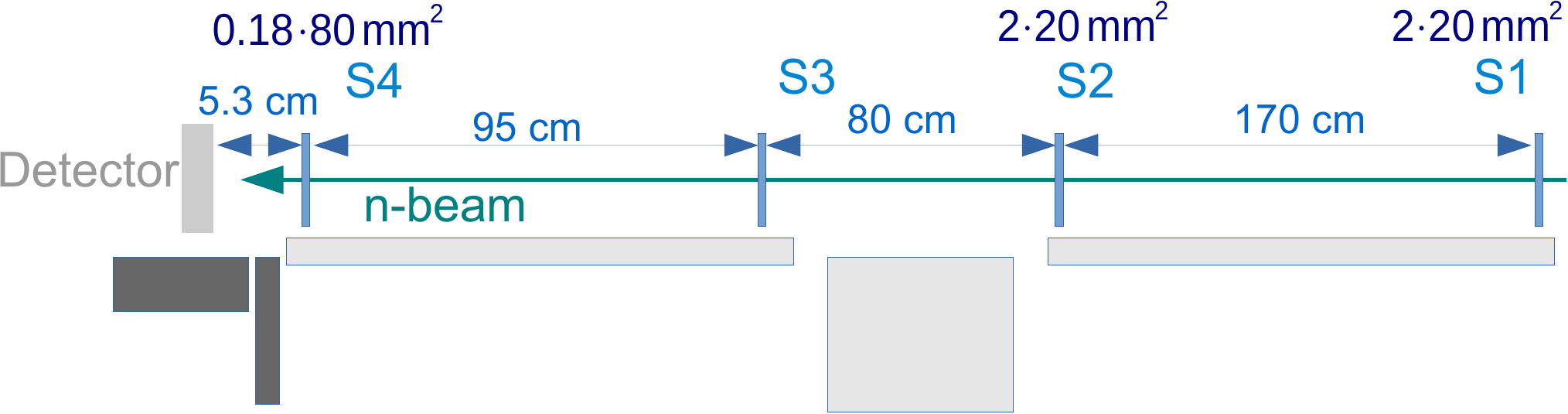}
\caption{Experimental setup for position resolution measurements at the Garching Research Reactor (FRMII) at the beam line TREFF. Neutrons of 4.7$\AA$ (13.4 meV) are collimated through a system of four apertures to a slit of 250 $\mu$m width and 65 mm height on the converter cathode of a GEM-detector}
\label{be}
\end{figure}

Neutrons of 4.7 \AA ~ are available at the Garching FRMII research reactor at the TREFF beam line. 
The beam was collimated by a system of four apertures, 
see figure \ref{be}. 
The collimation was done by the collimators 1, 2 and 4 and 
a possible beam halo was cut by collimator 3. 
The beam was formed to a slit of 180 $\mu$m  width and 65 mm height, 
defined by the collimators on the last aperture, 
which lead to a maximum width of the beam spot on the converter cathode 
of 250 $\mu$m  with only 0.32 mrad divergence.
To determine  the neutron profile on the conversion layer 
the hit distribution 
has been approximated by a boxcar function.
The convolution with the Gaussian-distribution 
describing the spatial detector resolution
allows then to simulate the measured distribution. 
This leads to a double error-function of the form:

\begin{equation}
f(x)= A \left(erf\left(\frac{a+\left(x-\mu \right)}{\sqrt{2}\sigma}\right)+erf\left(\frac{a-\left(x-\mu \right)}{\sqrt{2}\sigma} \right) \right)
\label{fit}
\end{equation}

$A$ denotes a normalization factor, $a$ is the half width of the slit, 
$\mu$ is the center of the slit and $\sigma$ is the standard deviation of the Gaussian, 
which represents the detector-resolution.

\begin{figure}
\centering
\includegraphics[width=\linewidth]{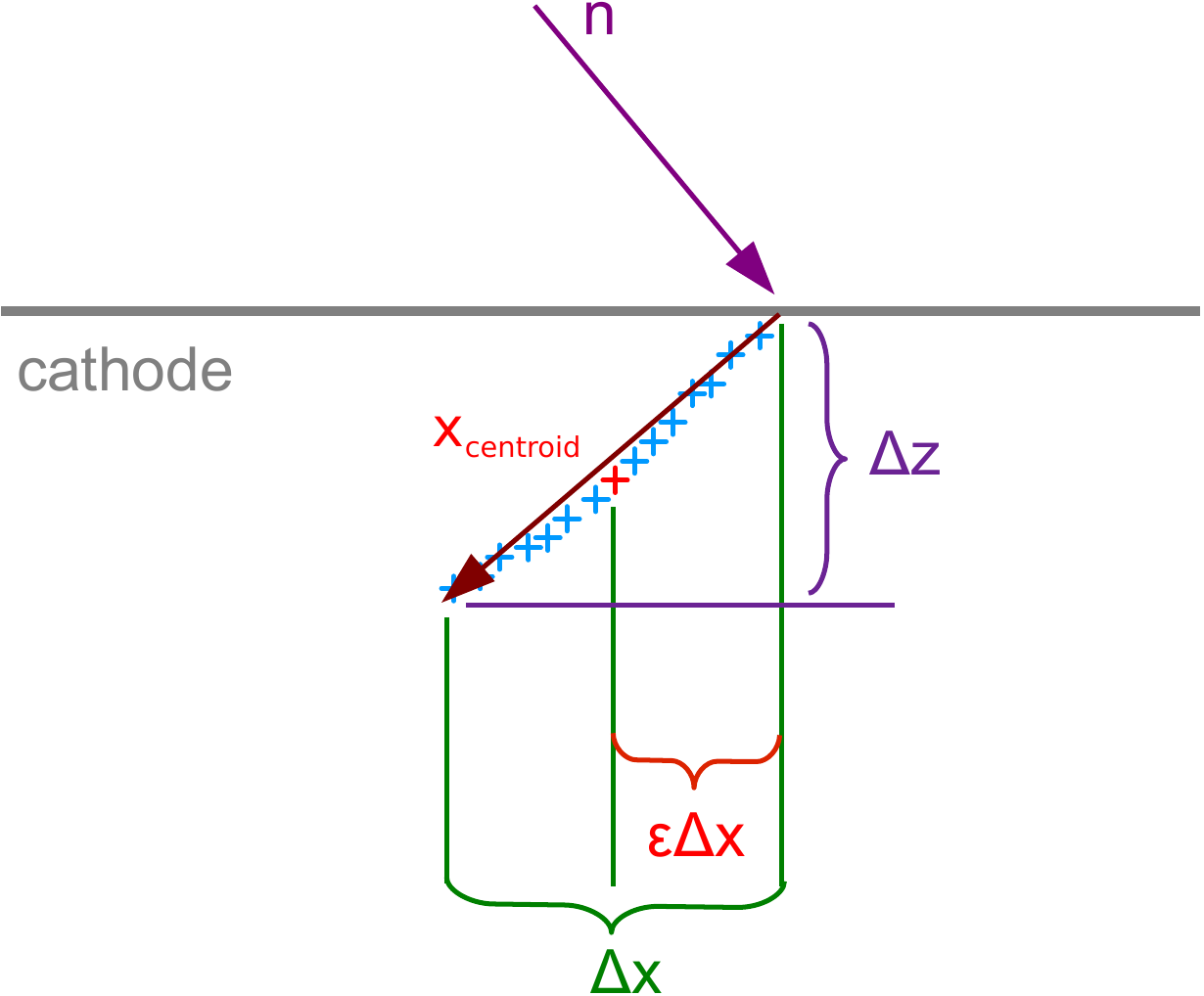}
\caption{The conversion point in the cathode can be determined 
starting from the center 
of charge position, which is well defined in the track, and 
applying a 
correction to the position, which is dependent on the 
direction and length of the track and the type of the ion producing the track}
\label{fig:conv-point}
\end{figure}

In a standard analysis the neutron position would be determined  by the center of the charge distribution 
for every track in the readout plane.
The error in position resolution is 
therefore limited by the inclination angle
and the length of the track of the secondary ion,
see figure \ref{fig:conv-point}.
The resolution achievable by this method is a FWHM of 3.4 $\pm$ 0.1 mm. 
\begin{figure}
\centering
\includegraphics[width=\linewidth]{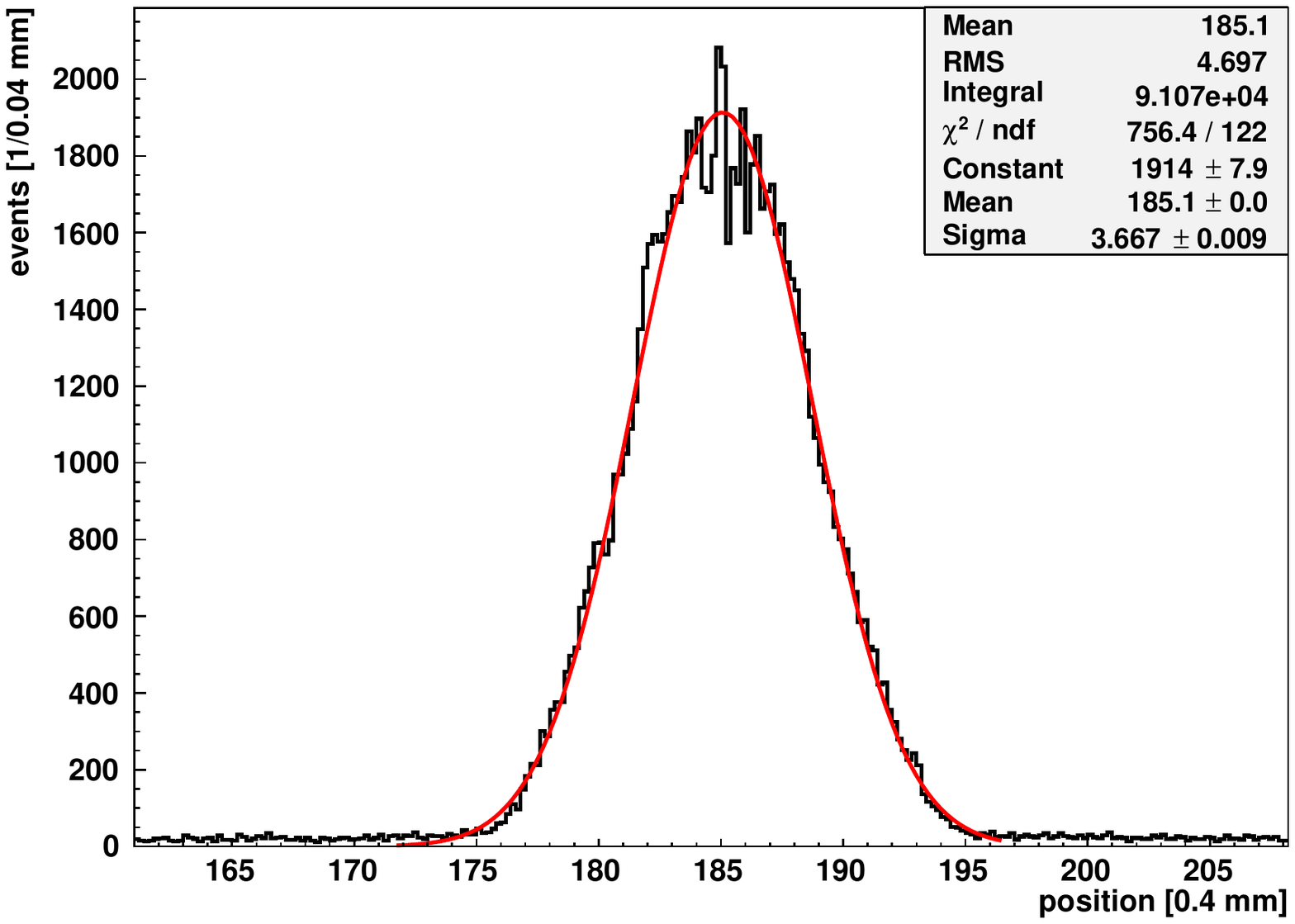}
                \caption{Projection onto the X-direction for charge averaged centers of tracks of a slit for a slit of 180 $\mu$m width. The width of the distribution is 3.4 mm (FWHM) ($\sigma=3.667 \times$0.4 mm ).  }
              \label{fig:ResoUn}
\end{figure}

The resolution can be improved by determining the starting point 
of the secondary ion track and thus the point of conversion in the cathode.

With a single GEM-detector spatial resolution of below 470\,$\mu$m could already be achieved by determination of the last strip, which was hit in a track(\cite{Pfeiffer}). 
In a triple GEM-setup the spread of the charge cloud by diffusion towards the anode is quite significant. This can be overcome by fully tracking the ions in the drift volume.

The method would work best if all ions would transverse 
the complete drift region and if the correlation between 
responding readout strips and ion track would be precise. 
Certainly, ions with large inclination angles 
stop within the 6 mm wide drift region. 
The systematic effects: electron diffusion and 
capacitive coupling of neighboring readout strips, 
lead to an overestimation of the length of the track
on both ends for both traversing and stopping ions. 
Therefore, the TPC-like reconstruction is 
dependent on the track length, the inclination angle, and 
the non-homogeneous the charge distribution in the track, the skewness.  
Skewness and track length are dependent on the particle type
as well. 

\begin{figure}
\centering
\includegraphics[width=0.9\linewidth]{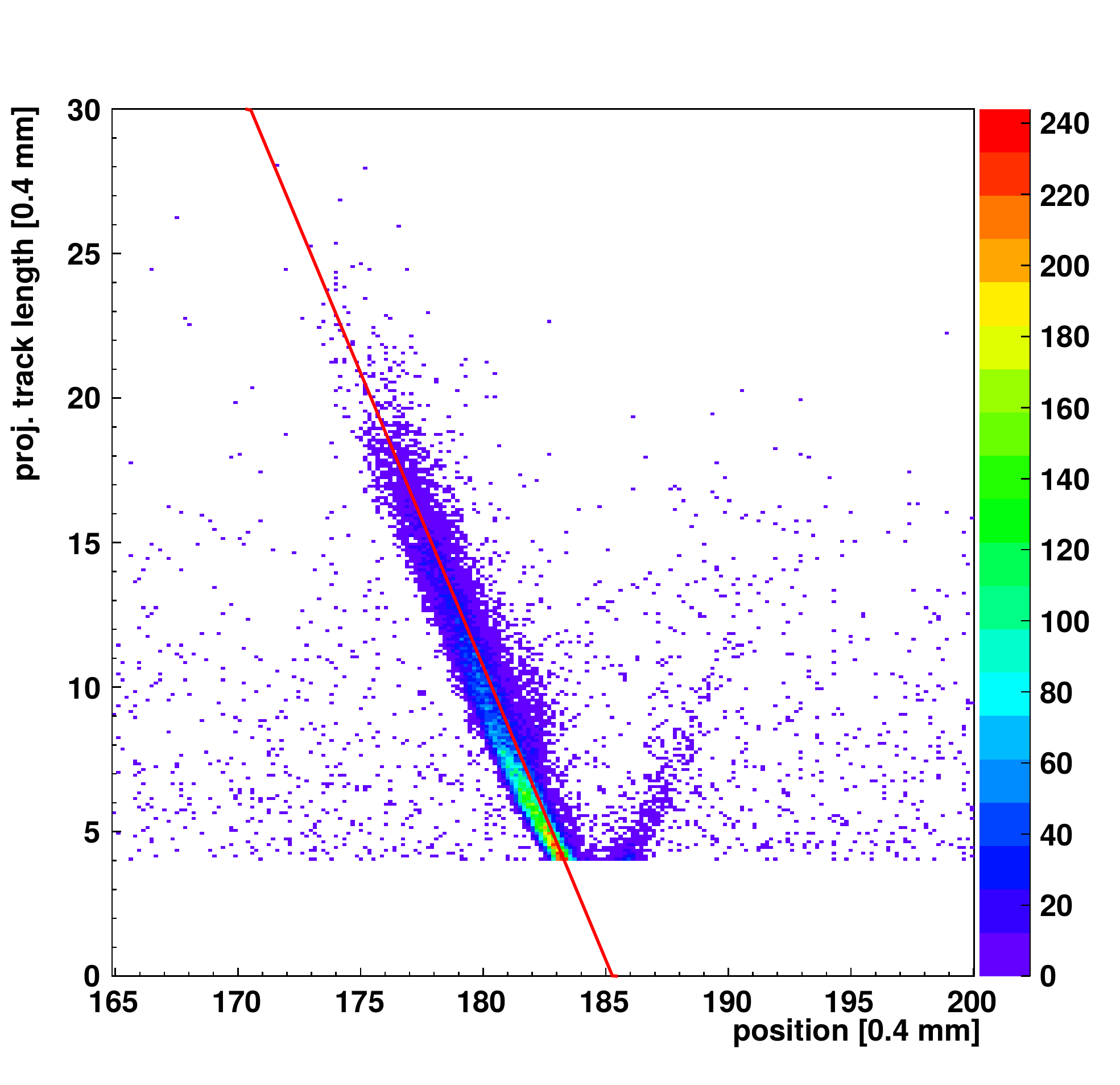}
\caption{Projected track length in X-readout direction plotted against the center of charge position for particles with negative inclination angle going from right to left in X-direction. A linear fit is applied to compensate for the track length dependence.}
\label{pic:linefit}
\end{figure}

\begin{figure}
\centering
\includegraphics[width=0.9\linewidth]{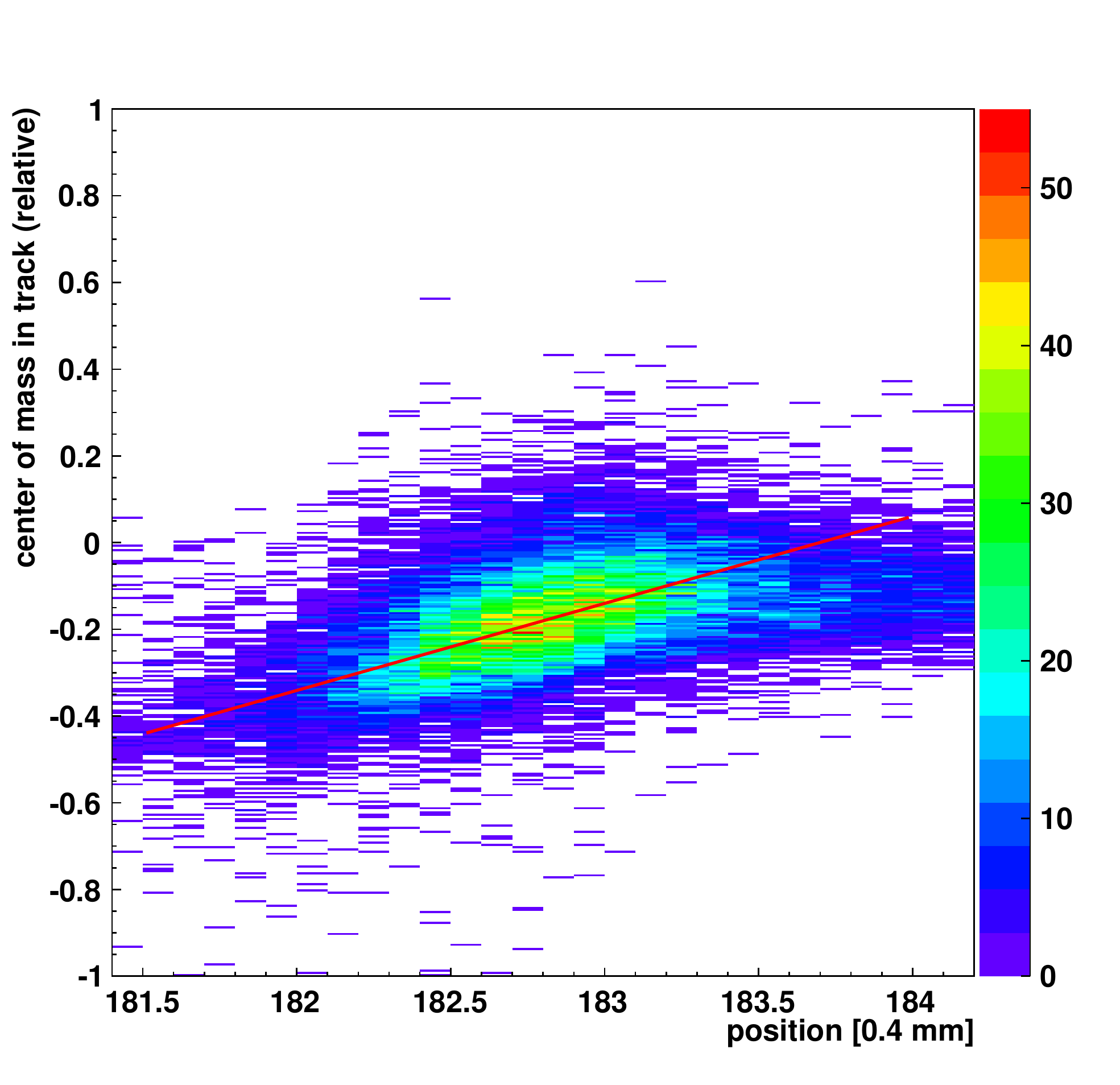}
\caption{Relative position of the center of charge in the track plotted against the track length corrected position. A center of charge position at the end of the track corresponds to 1 and (-1) to the begining of the track. The dependence is determined by a linear fit and the position is corrected for this dependence.}
\label{pic:etafit}
\end{figure}

\begin{figure}
\centering
\includegraphics[width=0.9\linewidth]{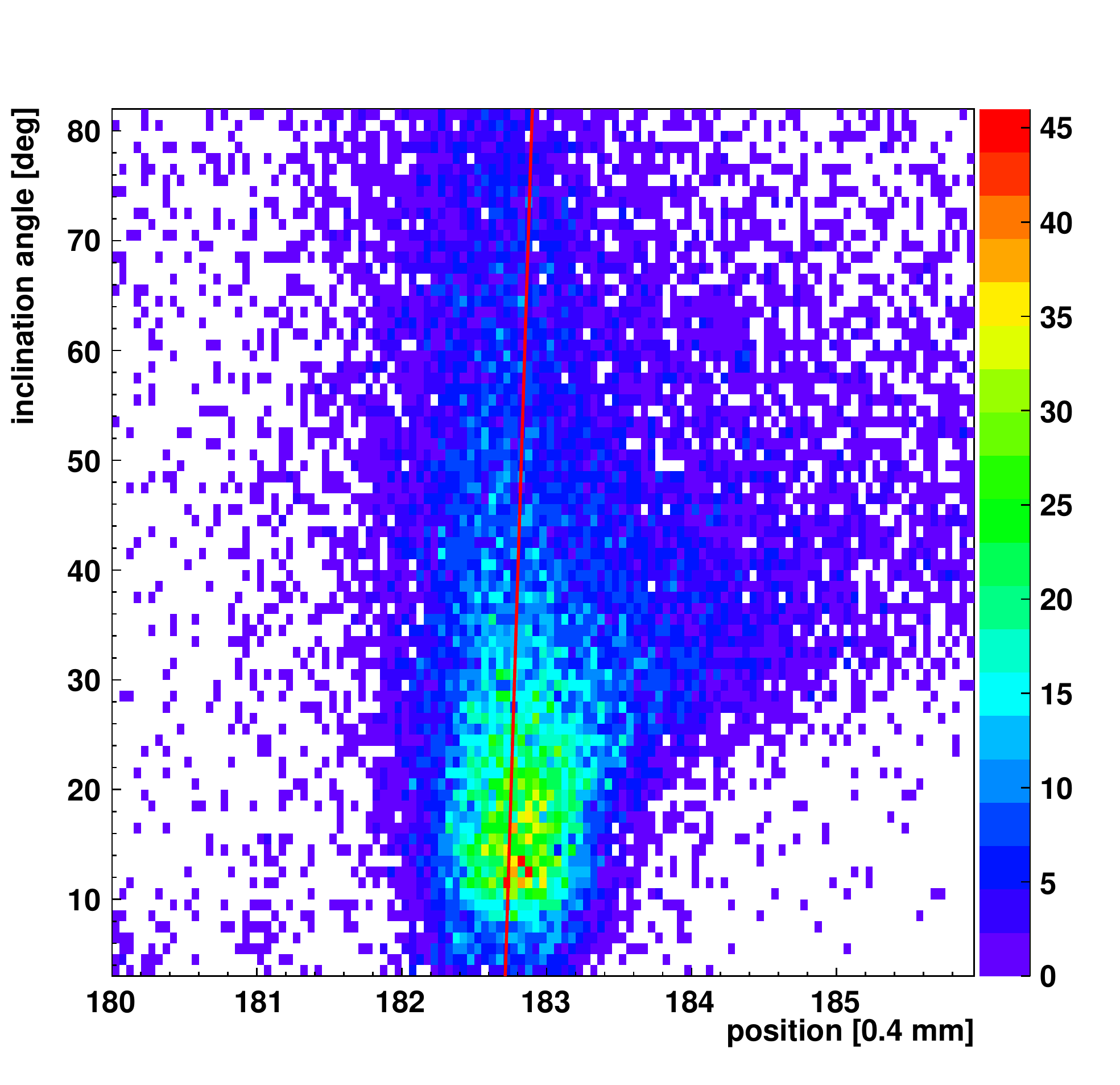}
\caption{Projected angle in X-readout direction plotted against corrected position. A linear fit is applied to compensate for the angular dependence.}
\label{pic:anglefit}
\end{figure}

In the following the dependencies of the reconstruction on the track length, 
and the skewness of the charge distribution in the track
will be discussed. 
These corrections are performed by linear  fits 
to internal detector correlations see figure \ref{fig:conv-point}.

Figure \ref{pic:linefit} shows a clear correlation between the projected track length in one readout direction of a track and the reconstructed position. The length of the track is determined from a Gaussian fit to the charge distribution on all responding strips individually in X and Y direction. 
The dependence of the centroid position on the projected track length was determined by linear fit and the position was corrected depending on the projected track length.

After application of this correction the dependence of the reconstructed position on
the skewness of the charge distribution is shown in figure \ref{pic:etafit}.
The linear fit gives the parameters for the correction
in a second step and in the last step the fit to the reconstructed angle of the track 
as a function of reconstructed position is investigated, as shown in figure \ref{pic:anglefit}. 

Since there are dependencies of all these measurement parameters it is not advantageous to correct them iteratively in a calibration one after the other.

%All together this approach delivers corrections of the form:
%\begin{align}
% \begin{split}
% x_0=x_{c}+ & \textrm{sign}(\left\theta\right)(0.84 \times \Delta x -0.0137\Delta x^2 + 0.008 %\cdot \theta + \\ + & 1.99 \cdot \eta +1.331 )
% \end{split}
%\end{align}
%Where x$_0$ is the reconstructed point of conversion, x$_c$ is the centroid position of the track, $\theta$ is the inclination angle $\Delta$x is the track length and $\eta$ is the skewness of the track.

\begin{figure}
\centering
                \includegraphics[width=\linewidth]{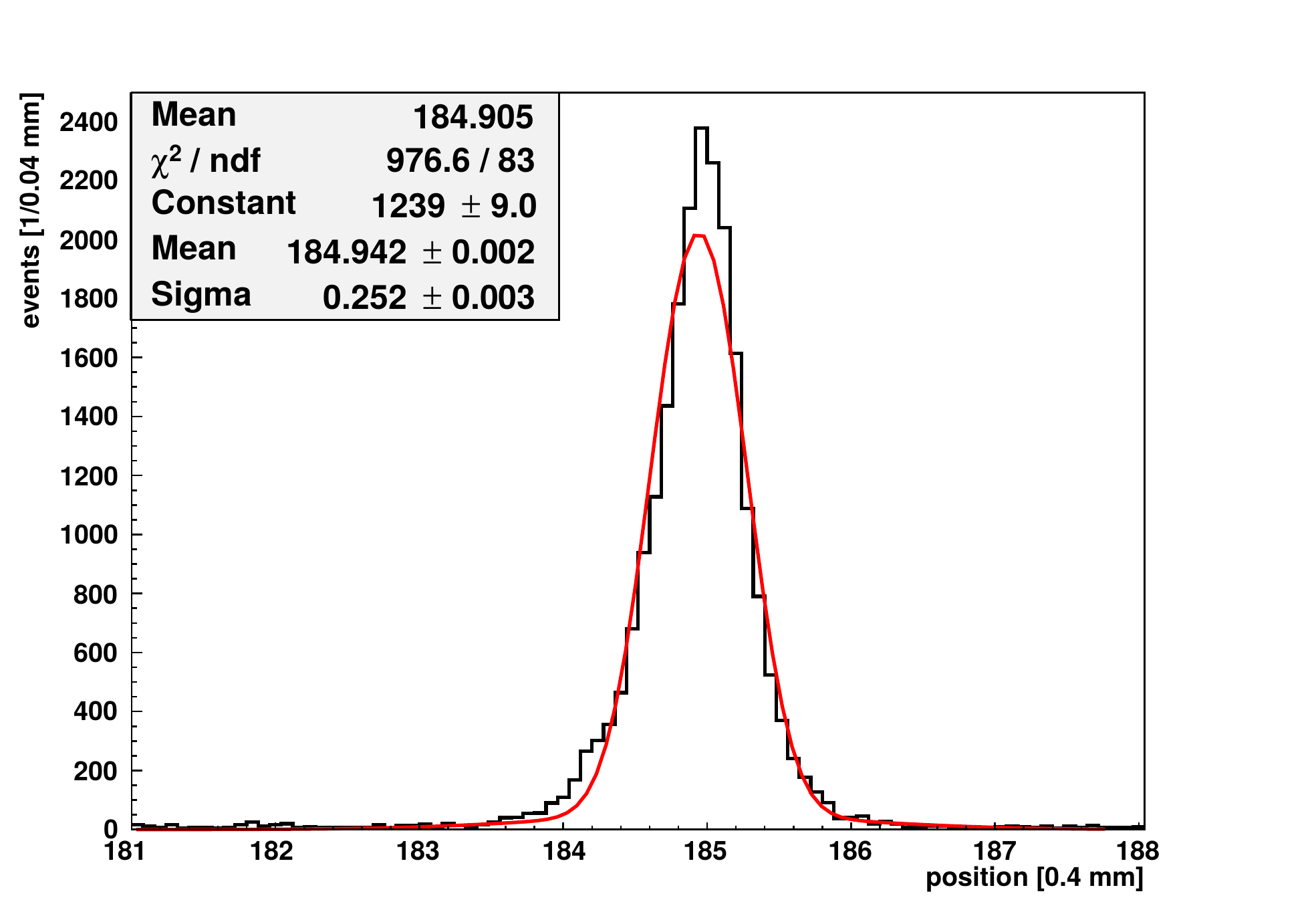}
                \caption{Hit pattern in one readout direction, after applying correction factors on track length, angle and charge distribution in the track. The achieved resolution is determined by a Gaussian function convoluted with a double error function and was determined to (235$\pm$ 20) $\mu$m FWHM  ($\sigma=0.252 \times$0.4 mm ) }
                \label{fig:X-Good}

\end{figure}

The result of this calibration is shown in figure \ref{fig:X-Good}, where the projection of the strip in the X-direction is shown and the resolution is determined by a fit of equation \ref{fit}. A spatial resolution of (235 $\pm$ 25) $\mu$m FWHM ($\sigma=$(100 $\pm$ 15) $\mu$m) is achieved.

\section{Conclusion}
 Using $\mu$TPC analysis the inclination angle of the track is determined, which defines the direction of the track.
As origin the centroid of the tracks was determined and its dependence on the track length, inclination angle,  and the skewness of the charge distribution in the track is examined. The iteration of calibration factors determined from linear fits to the reconstructed position in dependence to the mentioned track parameters were adapted to compensate for these.
Altogether this corrections in addition to a cut on the minimal track length of 3.5 strips in both directions led to a gaussian distribution with an improved resolution of (235$\pm$25) {$\mu$m} FWHM ($\sigma=$ {100$\pm$15} {$\mu$m}) in comparison to a standard centroid determination with a resolution of 3.4 mm. 

This procedure produces implicitly correction factors that have been so far obtained in MIP tracking experiments by dedicated calibration runs with known incidence angle of the particle beam.

\section*{Acknowledgments}
This work is based upon experiments performed at the TREFF instrument operated by FRMII at the Heinz Maier-Leibnitz Zentrum (MLZ), Garching, Germany.\\
The authors thank H. Takahashi (University of Tokyo) for providing of the $^{10}$B cathode.
%% bibliography

\end{document}